\documentclass[aps,prl,reprint,superscriptaddress, showpacs]{revtex4-1}

\usepackage[american]{babel}
\usepackage{amsmath,amsfonts,amssymb,epsfig}
\usepackage{float}
\usepackage{subfigure}
\usepackage{caption}
\usepackage{tikz}
\usepackage{graphicx}	%
\usepackage{epstopdf}	
\usepackage{xspace}		
\usepackage{gensymb}	
\usepackage{fancyhdr}
\usepackage[hidelinks]{hyperref}
\usepackage{multirow}
\usepackage{rotating}
\usepackage{color}
\usepackage[normalem]{ulem}
\usepackage{cancel}

\newcommand{\Angstrom}{\text{\normalfont\AA}\xspace}

\begin{document}
	\title{Self-generated plasma rotation in a Z-pinch implosion with preembedded axial magnetic field}
	
	\author{M. Cveji\'{c}}\email[correspoding author: ]{marko.cvejic@weizmann.ac.il}\affiliation{Weizmann Institute of Science, Rehovot 7610001, Israel}
	\author{D. Mikitchuk}\affiliation{Weizmann Institute of Science, Rehovot 7610001, Israel}\affiliation{Ecole polytechnique f\`{e}d\`{e}rale de Lausanne (EPFL), Route Cantonale, 1015 Lausanne, Switzerland}
	\author{E. Kroupp}\affiliation{Weizmann Institute of Science, Rehovot 7610001, Israel}
	\author{R. Doron}\affiliation{Weizmann Institute of Science, Rehovot 7610001, Israel}
	\author{P. Sharma}\affiliation{Weizmann Institute of Science, Rehovot 7610001, Israel}
	\author{Y. Maron}\affiliation{Weizmann Institute of Science, Rehovot 7610001, Israel}
	\author{A. L. Velikovich}
	\affiliation{Plasma Physics Division, Naval Research Laboratory, Washington, D.C. 20375, USA}
	\author{A. Fruchtman}\affiliation{Holon Institute of Technology, P.O. Box 305, Holon 58102, Israel}
	\author{I. E. Ochs}\affiliation{Department of Astrophysical Sciences, Princeton University, Princeton, New Jersey 08540, USA}
	\author{E. J. Kolmes}\affiliation{Department of Astrophysical Sciences, Princeton University, Princeton, New Jersey 08540, USA}
	\author{N. J. Fisch}\affiliation{Department of Astrophysical Sciences, Princeton University, Princeton, New Jersey 08540, USA}
	\date{\today}
	
\begin{abstract} \label{Abstract}
	Using detailed spectroscopic measurements, highly-resolved in both time and space, a self-generated plasma rotation is demonstrated for the first time in a cylindrical Z-pinch implosion with a pre-embedded axial magnetic field.  
	The plasma parameters and all three components of the magnetic field are resolved.  
	The plasma is seen to rotate at a velocity comparable to the peak implosion velocity, considerably affecting the force and energy balance throughout the implosion. 
	Moreover, the evolution of the rotation is consistent with magnetic flux surface isorotation, a novel observation in a Z-pinch. 

\end{abstract}

\maketitle

\label{IntroductionSection}
Z-pinch implosions with preembedded axial magnetic fields ($B_{z0}$) have been intensively studied during the last decade \cite{Slutz2010,Slutz2012,Awe2013,Awe2014,Sorokin2013,Qi2014,Mikitchuk2014,Gomez2014,Knapp2015,Atoyan2016,YagerElorriaga2016,Rousskikh2016,Rousskikh2017,YagerElorriaga2018,Mikitchuk2019,Conti2020,Campbell2020}
due to their importance both for inertial confinement fusion  \cite{Slutz2010,Slutz2012,Gomez2014,Knapp2015} and for the study of fundamental plasma physics \cite{Giuliani2015, Maron2020}. A primary difference from standard Z-pinches (without $B_{z0}$) is that during the implosion not only the plasma is compressed, but also the axial magnetic field, leading to a profound effect on the implosion dynamics. 
It was recently shown that the implosion dynamics is affected not only by the counter-pressure of the compressed $B_z$, but also by the current redistribution to a peripheral low-density plasma caused by the presence of $B_z$ \cite{Mikitchuk2019}, a process that was then also demonstrated in a numerical simulation \cite{Seyler2020}. 

Plasma rotation is ubiquitous both in space \cite{Kulsrud2010} and laboratory plasmas \cite{Gueroult2019, Lebedev2019}.
In wire-array Z-pinches, imploding plasmas have been rotated
by using a twisted wire-array load \cite{Ampleford2008} or generating a radial cusp field \cite{Bennett2015}.   
In contrast,
we report here, for the first time for a cylindrical Z-pinch implosion with $B_{z0}$, on a self-generated plasma rotation.
The observed rotation velocity is found to be comparable to the implosion velocity, and the rotation direction is seen to depend on the sign of $B_z$.
This rotation affects the implosion dynamics both by exerting a centrifugal force  and by mitigating plasma instabilities \cite{Turchi1976, Huneault2019, N.Rostoker1994, Velikovich1995}. 
The phenomena are investigated by unique simultaneous spectroscopic measurements of all three spatial components of the $B$-field, the ion radial and azimuthal velocities, and the plasma parameters, enabling examination of both the $B$-field distribution and rotation profile throughout the implosion. This information allows for a quantitative study of the rotation effect and the energy balance besides exploring the mechanism of the rotation.
	
According to a classic result of axisymmetric magnetohydrodynamics, in steady state, the plasma along each magnetic flux surface should rotate at the same angular frequency.
This so-called \emph{isorotation} is important in many steady-state laboratory devices such as Hall thrusters \cite{Smirnov2007,Fisch2011,Gao2016,Jiang2017,Liang2019}, centrifugal mirror fusion \cite{Bekhtenev1980,Volosov2006}, plasma mass filters \cite{Fetterman2011,Gueroult2014}, and in some models of accretion disc collapse \cite{Konigl1999,Salmeron2011,Bai2013,Bethune2017}.
However, the present measurements demonstrate, for the first time, rotation profiles consistent with a tendency toward isorotation in a Z-pinch implosion -- a distinctly non-steady-state system -- with the arrival towards isorotation occurring on the same time scale as the implosion.

\label{ExperimentSection}
\begin{figure}
	     \includegraphics[width=0.85\linewidth]{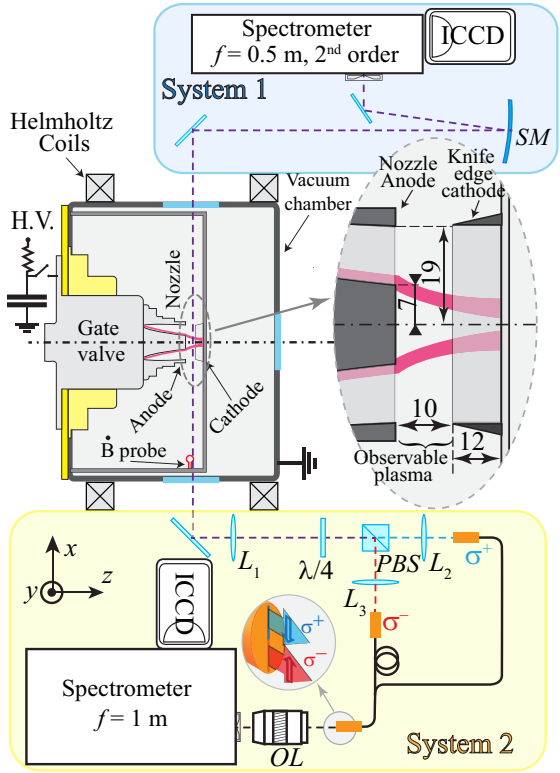}
	\caption{A schematic description of the experimental setup and spectroscopic system. The inset shows the geometry of the electrodes (mm) in the vicinity of the A-K gap. }
	\label{Fig1:ExpSeutp}
\end{figure}

Fig.~\ref{Fig1:ExpSeutp} shows schematics of the experimental and diagnostic setups. A cylindrical oxygen gas-puff shell
with initial outer and inner radii of 19 and 7 mm, respectively,
and mass $\sim 10$ $\mu$g/cm, as determined by interferometry, prefills the A-K gap with a pre-embedded, quasi-static axial magnetic flux ($B_{z0} = 0.26~$ T).
$B_{z0}$ is generated by a pair of Helmholtz coils (HC), placed outside of the vacuum chamber, carrying a long current pulse ($\sim 95$ ms) to allow for the diffusion of $B_{z0}$ into the vacuum chamber and the A-K gap. 
Subsequently, a pulsed current (rising to 300 kA in 1.6 $\mu$s) is driven through the gas, ionizes it, and generates an azimuthal magnetic field that compresses the plasma radially inward together with the embedded $B_z$ field. 
The time of the peak plasma compression (denoted in this paper as $t = 0$) is determined using the peak of a UV-VIS photo-diode signal that coincides with the minimum plasma radius seen in 2D images, occurs for $B_{z0} = 0.26$ T,  $\sim700$ ns after current initiation. 
A $B$-dot probe measures the total current. 

The diagnostic setup is designed for acquiring spatially and temporally resolved data for simultaneous determination of the plasma parameters, the $B$-fields, and ion motion.
It consists of two imaging spectroscopic systems; their main parameters are given in Table \ref{Table_SpectroscopyParam}. 
The line-of-sights of both systems are perpendicular to the pinch axis. 
Compressing azimuthal ($B_\theta$) magnetic field is measured simultaneously with axial ($B_z$) and radial ($B_r$) components of the compressed magnetic field. 
The radial distribution of the ion velocities, $B$-field and plasma parameters are obtained by taking advantage of the radial separation of ion charge states during implosion, without the need to use Abel inversion \cite{Davara1998, Rosenzweig2020,Rosenzweig2017,Maron2020}.

\begin{table}
	\caption{Summary of the spectroscopic-system parameters}
		\begin{ruledtabular}
			\begin{tabular}{ccc}
				Parameter & System 1 & System 2 \\
				\hline
				Spectral resolution & $0.7~\Angstrom$ at $5575~ \Angstrom$, & $0.3~ \Angstrom$ at $3800~\Angstrom$  \\
				Spatial resolution & $<0.2$ mm & $0.7$ mm \\
				Typical ICCD gate time & $\sim 5$ ns & $\sim 30$ ns \\
				Minimal detectable velocity &$0.7 \cdot 10^{4}$ m/s &$0.3\cdot 10^{4}$ m/s 	 	
			\end{tabular}
		\end{ruledtabular}
	\label{Table_SpectroscopyParam}	
\end{table}

\begin{figure}
	\includegraphics[width=0.85\linewidth]{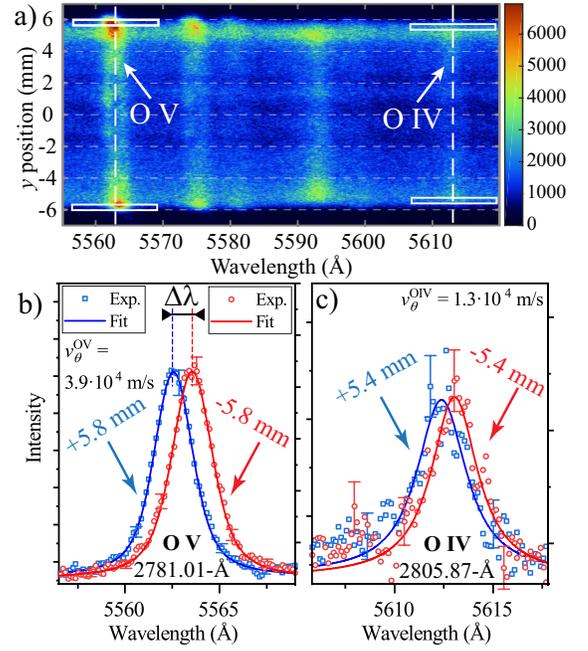}
	\caption{a) A spectral image of chordally integrated oxygen plasma emission recorded at $t= -63$ ns and $z = 9$ mm. b) Lineouts of the OV transition at $\lambda= 2781~ \Angstrom$ (recorded in the $2^\text{nd}$ spectral order) representing emission at $r = 5.8$ mm (blue squares) and $r = -5.8$ mm (red circles), together with the best fit to the shape of each line using Voigt profile; c) Similar to (b) for an OIV transition at $\lambda = 2806~\Angstrom$. The lineouts are generated from the regions of the spectral image marked by rectangles. Unperturbed line centers are marked with dashed lines.}
	\label{Fig2:RotationSpectra}
\end{figure}

\label{Results}
Fig.~\ref{Fig2:RotationSpectra} shows a typical spectral image of system 1 recorded at $t = - 63$ ns (63 ns before the peak of the photo-diode) and $z = 9$ mm. 
The most prominent lines are: O V 
at $\lambda=2781~\Angstrom$, O IV at $\lambda=2806~\Angstrom$, and O III at $\lambda = 5592~\Angstrom$.  
The rotation velocities are determined from the Doppler-shifted line emissions at $y = \pm r^{\text{O V}}, \pm r^{\text{O IV}}$, where $r^{\text{O V}}$ and $r^{\text{O IV}}$ are the outer radii of the O IV and OV emissions, defined at the $r$-position of $50\%$ of the peak emission as it drops towards the larger radius (see the white rectangles in Fig. \ref{Fig2:RotationSpectra}a). Additional details of the rotation velocity measurements are given in the supplementary material \cite{SM}.
The rotation velocity is given by $v_\theta = \frac{1}{2}c\frac{\Delta \lambda}{\lambda_0}$, where $c$ is the speed of light, $\lambda_0$ is the transition wavelength, and $\Delta \lambda$ is the difference between the Doppler shifted line-centers for $+ r$ and $- r$ (Fig. \ref{Fig2:RotationSpectra}b). 
Similarly, the implosion velocity ($v_r$) of each charge state is obtained using lineouts generated from the $y=0$ chords of the spectral image. 
It was verified that the Stark shifts of the recorded spectral lines have a negligible effect on the ion velocity determination. 
The electron density $n_e$ is determined from the Stark broadening of the spectral lines \cite{stark-b}. 
The electron temperature, $T_e$, is obtained using line-intensity ratios of O V and O IV transitions, using a collisional-radiative model \cite{Ralchenko2001}. 
The evolution of the plasma parameters will be discussed in a separate study.

Spectroscopic system 2 recorded (simultaneously with system 1) the O VI and O III transitions at $\lambda = 3811~\Angstrom$ and $\lambda = 3791~\Angstrom$, respectively, which were used to determine the azimuthal ($B_\theta$), axial ($B_z$), and radial ($B_r$) magnetic fields, and for the measurements of $v_r$ and $v_\theta$ as it is done by system 1. 
The $B$-field determination is based on the polarization properties of the Zeeman effect. 
Description of the different spectroscopic methods based on the Zeeman effect for $B$-field measurements can be found in \cite{Doron2014}, while a detailed description of the system-2 setup and the data analysis of the $B_\theta$ measurement is given in \cite{Mikitchuk2019,Rosenzweig2020}, of the $B_z$ measurements in \cite{mikitchuk:2019b}, and of the $B_r$ measurements in the supplementary material \cite{SM}. 

\label{ResultsSection}

\begin{figure*}
	\centering
		\includegraphics[width=0.90\textwidth]{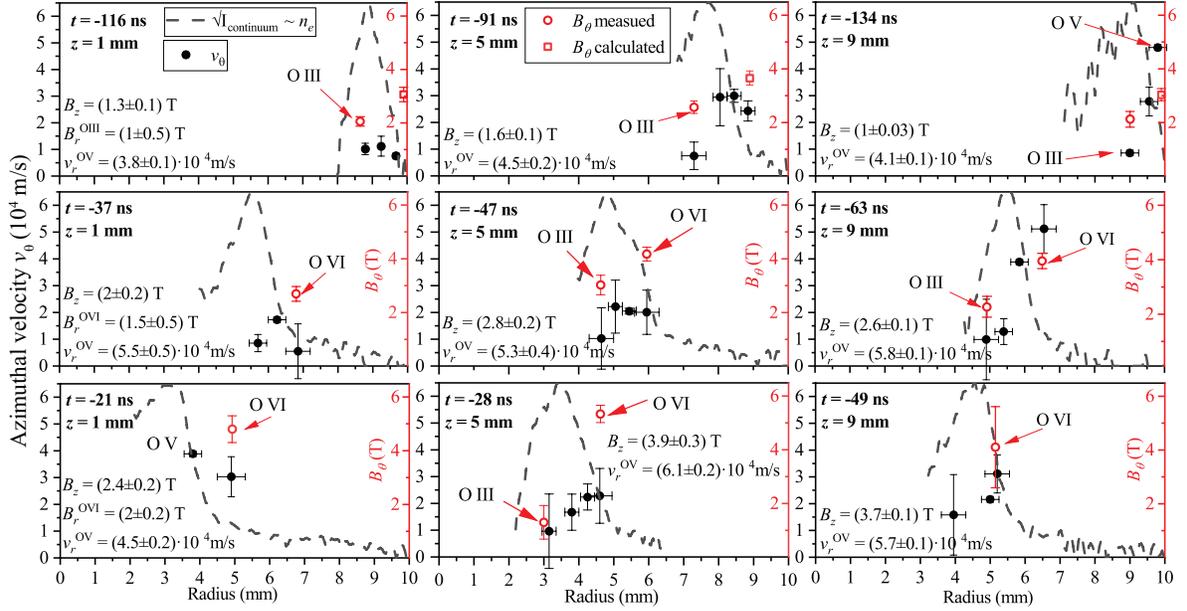}
	\caption{
		Radial distribution of $v_\theta$ (black full circles) and $B_\theta$ (red hollow circles) for different times and $z$ positions. The dashed line represents the square root of the radial distribution of continuum intensity.
	}
	\label{Fig3:RotResults}
\end{figure*}

The plasma velocities and magnetic field measurements are summarized in Fig.~\ref{Fig3:RotResults}, that shows the radial distribution of $v_\theta$ for different times and $z$-positions, together with the $B_\theta$ distribution. 
The dashed lines represent the radial distribution of the square root of the continuum emission averaged over $5522\pm4~\Angstrom$. 
For the relevant $T_e$ range ($7-13$ eV) it is nearly proportional to the electron density \cite{Griem1964}. 

In addition, each graph in Fig.~\ref{Fig3:RotResults} shows the $B_z$ value on axis and $v_r$ of the O V ions (for comparison with $v_\theta$). 
$B_r$ is given for $z = 1$ mm, since at $z=5$ and 9 mm it was too low to be detected.
We emphasize that the data presented in each graph is obtained in a single shot, while those given in different graphs are obtained in different shots. 

The measured values of $B_\theta$, $B_z$, $B_r$, $v_\theta$, $v_r$, $n_e$, and $T_e$, presented in Fig. \ref{Fig3:RotResults}, are essential for the evaluation of the forces acting on the plasma during implosion and for the understanding of the mechanisms responsible for plasma rotation.
In Table \ref{Table_MeasuredValues} we compare the different terms of the radial component of the equation of motion calculated at the outer radial position of O V:
\begin{equation}
{
	\rho \frac{dv_r}{dt} = - \frac{\partial p}{\partial r} - j_z B_\theta + j_\theta B_z + \rho\frac{v_\theta^2}{r}
}
\label{EquationOfMotion}
\end{equation}
where $p = (1 + \frac{1}{\bar{Z}}) n_e k_B T_e$ is the thermal pressure, $j_z \approx \frac{B_\theta}{\mu_0\Delta r}$ and $j_\theta \approx \frac{B_z}{\mu_0\Delta r}$ are the axial and azimuthal current densities, respectively, $\rho$ is the plasma mass density, and $\Delta r$ is the radial scale over which the plasma and $B$-field parameters change. 
$\Delta r$ is estimated using the $B_\theta$ measurements (see the plot for $z = 5$ mm, $t = - 28$ ns in Fig. \ref{Fig3:RotResults}) and the radial distribution of the continuum emission (see in Fig. \ref{Fig3:RotResults}).

Calculations presented in Table II show that the centrifugal force $f_C = \rho\frac{v_\theta^2}{r}$ plays a significant role in the plasma dynamics during the implosion. For example, at $z = 9$ mm and $t = 134$ ns, $f_C$ is the dominant outward acting force, i.e., $f_C > f_{B_z} + f_T$, where $f_{B_z} = j_\theta B_z$, $f_T = \frac{n k_B T}{\Delta r}$, $f_C$ is also $\approx 50\%$ of the compressing force $f_{B_\theta} = j_z B_\theta$.

\begin{table*}
	\caption{Calculation of different terms in Eq. \ref{EquationOfMotion}: $f_{B_\theta} = j_z B_\theta$,  $f_{B_z}=j_\theta B_z \approx \frac{B_z^2}{2\mu_0 \Delta r}$,$f_T = \frac{nk_B T}{\Delta r}$,$f_C = \rho\frac{v_{\theta}^2}{r}$}
	\begin{ruledtabular}
		\begin{tabular}{cccccccccccccc}
			$z-$position &Time &$r$ &$B_\theta$ &$B_z$ &$j_z$           &$\rho$       & $v_\theta$ &$T_e$ &$n_e$             &$f_{B_\theta}$  &$f_{B_z}$  &$f_T$  &$f_C$ \\
			(mm)     &(ns) &(mm)&(T)        &(T)   &($10^9$ Am$^{-2}$)&g/m$^{3}$   &10$^4$ m/s  &(eV)  &10$^{24}$ m$^{-3}$&\multicolumn{4}{c}{($10^9$ Nm$^{-3}$)} \\
			\hline 
			5        &-91  &8.7 &3.6        &1.6   &1.5             &7.6          &3           &9     &1                 &5.4             &0.5        &0.9    &0.8 \\
			5        &-28  &4.3 &4.4        &3.9   &2               &13.7         &2.2         &11    &2                 &8.8             &3          &2.2    &1.6 \\
			9        &-134 &9.8 &3          &1     &1.2             &7.8          &4.8         &9     &1                 &3.6             &0.2        &0.9    &1.8 \\
			9        &-49  &5   &4          &3.7   &1.7             &13.7         &2.2         &11    &2                 &6.8             &2.7        &2.2    &1.3
		\end{tabular}
	\end{ruledtabular}
	\label{Table_MeasuredValues}
\end{table*}

Additionally, the rotation might affect the plasma dynamics by the mitigation of MRT and MHD instabilities, either by affecting the effective acceleration direction  
of the Z-pinch \cite{N.Rostoker1994, Velikovich1995,Turchi1976, Huneault2019}, or by phase mixing of an instability perturbation due to the non-uniform angular velocity ($\omega = v_\theta / r$) distribution \cite{Shumlak1995,Shumlak2001}. 
Indeed, in almost all Z-pinch experiments with preembedded $B_z$, significant instability mitigation was observed during the implosion and stagnation \cite{Mikitchuk2014,Awe2013,Awe2014,Rousskikh2016, Rousskikh2017, Felber1988, Felber1988a, Qi2014, Conti2020}. 
It is a common assumption that the observed stabilization is solely due to the radial bending of the $B_z$-field lines \cite{Budko1989}. 
However, it is possible that also self-generated rotation, such as observed here, is partially responsible for the observed stabilization. 
For example, adopting the stabilization condition for the kink instability given in \cite{Shumlak1995} due to the sheared plasma velocity, and applying it to our case accounting for the non-uniform $\omega$, yields the condition $r\cdot\frac{\Delta\omega}{\Delta r}>10^7 \text{ s}^{-1}$ (see supplementary material \cite{SM}). 
Analyzing the $v_\theta$ distribution at $z=5$ mm and $t = -28$ ns (Fig. \ref{Fig3:RotResults}), we obtain $r\frac{\Delta\omega}{\Delta r} \approx 7\cdot 10^6 \text{ s}^{-1}$, which is close to the condition for stabilization.

\label{DiscussionSection}	
	
	Fig. \ref{Fig3:RotResults} shows several trends in the evolution of $v_\theta$.
	Early in the implosion, the plasma rotates fast near the cathode. The implosion then progresses,  with axial	asymmetry, where the plasma reaching a smaller radius near the cathode.
	Simultaneously, the rotation near the cathode slows down, while the rotation near the anode speeds up,
	reducing the difference between the rotation speeds at the different parts of the plasma.
	
	This unintuitive evolution of the rotation profile likely results from the fact that MHD systems tend to rotate at the same frequency \cite{Lehnert1971}, as outlined in Ferraro's isorotation theorem \cite{Ferraro1937,Kulsrud2010,Alfven1963}.
	This theorem says that in steady state for an axisymmetric MHD system:
	\begin{align}
		\nabla \omega \cdot \mathbf{B}_{p} = 0,
	\end{align}
	where $ \omega $ is the angular rotation frequency and $\mathbf{B}_p$ is the poloidal ($r$-$z$) magnetic field.
	The axial asymmetry in the implosion results from flux freezing in the magnetic nozzle (the anode side), which causes the flux surfaces to bend inward as they approach the anode.
	Isorotation slows the rotation near the cathode, and speeds it near the anode (Fig. \ref{Fig4:geometry}b).
	As we show in the supplementary material \cite{SM}, this mechanism can modify the rotation velocity by a factor of $\mathcal{O}(1)$, when the axial magnetic field is large enough for a shear Alfven wave to propagate across the anode-cathode gap over the implosion time.

\begin{figure}[t]
	\includegraphics[width=1\linewidth]{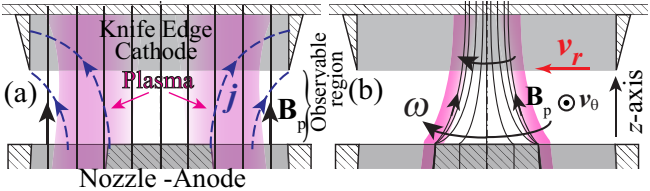}
	\caption{Schematic of the plasma, electrodes, velocities and poloidal $B$-field configuration at a) $t \approx -300$ ns and b) $ t \approx -80$ ns. $\omega = v_\theta / r$ is the rotation velocity and $\mathbf{B}_{p} = B_z \hat{z} + B_r\hat{r}$ is the magnetic field in the poloidal plane.
	}
	\label{Fig4:geometry}
\end{figure}
	
	The initial rotation of the plasma cannot be understood only based on MHD, since it depends on the initial current breakdown that is determined by the electrode geometry and initial gas distribution.
	While the anode is distributed evenly over large and small radius, the \textquotedblleft knife edge\textquotedblright~  round cathode is located at large radius.
	Thus, we expect a radial component to the current to occur near the cathode, which increases close to the cathode (Fig. 4a).
	This radial current combines with the $B_z$ to produce a $\mathbf{j} \times \mathbf{B}$ force in the $v_\theta$ direction (note that by our convention, the coordinate system $(r,z,\theta)$ is right-handed).
	Torque due to the $\mathbf{j} \times \mathbf{B}$ force initiates rotation near the cathode and causes a faster rotation at radii with lower plasma density (thus lower inertia) at the outer edge of the plasma, as observed (see Fig. \ref{Fig3:RotResults}, for $z=9$ mm).
	Using a simple snowplow model, we show that a radial current $I_r \sim 0.2 I_{total}$ is sufficient to bring the plasma near the cathode to $v_{\theta} \approx 4.8 \cdot 10^4$ m/s at $t=-134$ ns (see supplementary material \cite{SM}).
	We note that the line tying effect at the anode, see Fig. \ref{Fig4:geometry}b, might also affect the plasma dynamics near the anode by suppressing rotation early in the implosion. 

	In summary, novel spectroscopic measurements demonstrated self-generated plasma rotation for the first time in an imploding Z-pinch with an embedded axial magnetic field.  
	The rotation was shown to have two important features: One, the rotation velocity was comparable to the implosion velocity.  
	Two, the plasma rotation velocity profile was remarkably consistent with isorotation of the magnetic surfaces, even as the plasma underwent implosion.
	
	The fact that the rotation velocities were comparable to the implosion velocities indicates that the associated centrifugal forces exert a significant influence on force and energy balance. 
	Moreover, the large shear in the rotation suggests that it might exert a stabilizing effect on plasma instabilities, an effect consistent with observations in other rotating plasma experiments.

	The fact that the rotation is consistent with isorotation suggests that magnetized Z-pinch configurations designed for fusion purposes might also be useful in simulating other phenomena. 
	In particular, the observed rotation transport and magnetic geometry bear marked similarity to the magneto-centrifugal wind model of stellar accretion disc collapse \cite{Konigl1999, Salmeron2011, Bai2013, Bethune2017}, where isorotation leads to angular momentum transport outward along the flux surfaces, allowing the disc to collapse and the star to form. 
	While the connection to accretion discs is somewhat speculative at this stage, it does suggest that study of rotation in laboratory Z-pinches might shed light on rotating systems for which the set of detailed measurements as performed here would be impossible.

\label{Acknowledgments}	
\begin{acknowledgments}
	We thank A. Fisher and U. Shumlak for invaluable
	suggestions and fruitful discussions and P. Meiri for his skillful
	assistance. This work was supported in part by the Cornell Multi-University Center for High Energy Density Science (USA), by Lawrence Livermore National Laboratory, by the USA-Israel Binational Science Foundation, by NSF-BSF (USA-Israel), by  NNSA 83228-10966 [Prime No. DOE (NNSA) DE-NA0003764] and by NSF PHY-1805316. 
\end{acknowledgments}
M.C. and D.M. contributed equally to this work.


%

\end{document}